\documentclass[preprint,showpacs,preprintnumbers,amsmath,amssymb]{revtex4}

\usepackage{graphicx}
\usepackage{dcolumn}
\usepackage{bm}
\usepackage{color}

\begin{document}

\preprint{APS/123-QED}

\title{Spin Relaxation Mechanism in a Highly Doped Organic Polymer Film}

\author{Motoi Kimata$^1$*, Daisuke Nozaki$^1$, Yasuhiro Niimi$^1$, Hiroyuki Tajima$^2$, YoshiChika Otani$^{1, 3}$}
\email{kimata@issp.u-tokyo.ac.jp}
\affiliation{$^1$Institute for Solid State Physics, University of Tokyo, Kashiwano-ha 5-1-5, Kashiwa, Chiba, Japan 277-8501 \\
$^2$Graduate School of Material Science, University of Hyogo, Kouto 3-2-1, Kamigori, Ako, Hyogo, Japan 678-1205 \\
$^3$Center for Emergent Matter Science (CEMS) RIKEN, 2-1 Hirosawa, Wako, Saitama 351-0198, Japan }%

\date{\today}

\begin{abstract}
We report the systematic studies of spin current transport and relaxation mechanism in highly doped organic polymer film. In this study, we have determined spin diffusion length (SDL), spin lifetime, and spin diffusion constant by using different experimental techniques. The spin lifetime estimated from the electron paramagnetic resonance experiment is much shorter than the previous expectation beyond the experimental ambiguity. This suggests that significantly large spin diffusion constant, which is reasonably explained by the hopping transport mechanism in degenerate semiconductors, exists in highly doped organic semiconductors. The calculated SDL using the spin lifetime and spin diffusion constant estimated from our experiment is comparable to the experimentally obtained SDL of the order of one hundred nanometers. Moreover, the present study revealed that the spin angular momentum is almost preserved in the hopping events. In other words, the spin relaxation mainly occurs due to the spin-orbit coupling at the nanoscale crystalline grains. 

\end{abstract}

 
\maketitle

\section{Introduction}

Organic semiconductors (OSCs) have been attracting much attention because of their potentiality to be applied for field-effect transistors, solar cells, and displays. These application studies are stimulated by the unique properties of OSCs, e.g., low-cost, low-weight, and flexibility. In addition to such unique characters, OSCs are attractive candidates for spintronic device application owing to following reasons. The spintronics use electron spins as information carriers, and therefore their lifetime and transport length, i.e., spin diffusion length (SDL), are important parameters to design spintronics devices \cite{zutic}. OSCs generally consist of relatively light elements, such as hydrogen, carbon, and sulfur of which spin-orbit (SO) interaction is expected to be weak. Since the SO interaction is a main origin of the spin relaxation, OSCs is a promising material for long-distance spin transport \cite{dediu}.  
So far, many experiments have been carried out to measure the SDLs of OSCs by using magnetoresistance effect between two ferromagnetic electrodes, where the spin polarized charge currents pass through the organic layer \cite{dediu2, xiong, morley, santos, nguyen, shim}. On the other hand, the pure spin current accompanying no net charge current, is an essential ingredient of next generation spintronics \cite{fukuma, idzuchi}. However, the study of the pure spin transport in OSCs is very limited, and their properties are not yet fully understood because of the lack of complementary information about the spin diffusion constant ($D_{\rm S}$) and the spin lifetime ($\tau_{\rm S}$). For example, recently-performed dynamical spin transport experiments suggest quite long $\tau_{\rm S}$ in OSCs, but their reliability has not been confirmed experimentally \cite{ando, watanabe}. To elucidate this problem, the comprehensive study of OSCs to determine spin transport parameters is necessary. 

\begin{figure}
\includegraphics[width=5.5cm]{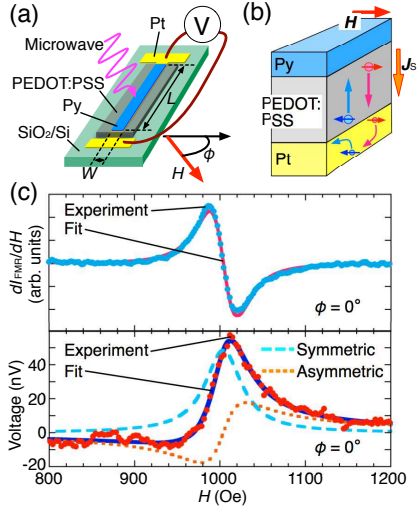}
\caption{\label{f1}(color online). (a)Schematic of the sample structure used for the spin transport experiment and (b)the mechanism of spin injection and detection. 
The injected pure spin current through the PEDOT:PSS layer is absorbed by the Pt layer, and then converted to the electric field via the ISH effect. (c) (upper panel)The FMR spectra of the Py strip of Py(17)/PEDOT:PSS(60)/Pt(8) trilayer sample. 
The numbers in parentheses indicate the thickness in nanometers. 
(lower panel)The dc voltage signal at the Pt layer of the trilayer sample for $\phi=0^{\circ}$. }
\end{figure}

In the present work, we have performed the comprehensive studies on the spin transport by means of spin pumping, electron paramagnetic resonance (EPR), and charge transport experiments using a highly doped organic polymer film. From these experiments, we have determined the SDL, $\tau_{\rm S}$, and $D_{\rm S}$ independently, and then deduced the spin transport mechanism in disordered organic polymers. 

The organic material focused in this study is a conducting polymer poly(3,4-ethylenedioxythiophene): polystyrene sulfonate (PEDOT:PSS, PH1000 Clevios). In this material, the conjugated PEDOT polymer is doped with PSS. The morphology and synchrotron X-ray diffraction studies reveal the structure of the PEDOT:PSS film, in which the pancake-like core-shell structure constructed by a PEDOT rich core and a PSS insulating shell is assembled \cite{nardes, takano}. The dimensions of this core-shell structure are 20 - 30 nm and 5 - 6 nm along the in-plane and out-of-plane directions, respectively. 

\section{Experimental details}

The device structure used in the dynamical spin transport experiment is a Ni$_{81}$Fe$_{19}$ (Permalloy: Py)/PEDOT:PSS/Pt trilayer. The device is fabricated on a thermally oxidized Si substrate (thickness of SiO$_2$ is 100 nm). After the deposition of the Pt layer, a water distributed solution of PEDOT:PSS is spin coated with rotational speeds of 1000 - 6000 rpm to change the thickness of PEDOT:PSS layer. The surface roughness of the PEDOT:PSS film is evaluated as approximately 3 - 4 nm by an atomic force microscopy. The PEDOT:PSS film is annealed at 50$^{\circ}$C in a high vacuum about 10 hours, and then the Py layer is deposited on PEDOT:PSS at a rate of $\sim$0.1 ${\rm \AA}$/s. The spin pumping and EPR experiments were performed by using a conventional X-band EPR spectrometer with a TE$_{102}$ rectangular cavity. The operation frequency is approximately 9.45 GHz, and the sample is located near the center of the cavity. The charge transport measurements were performed in the same trilayer structure with a junction area of 40$\times$100 ($\mu$m)$^2$.

\section{Results and discussions}

\begin{figure*}
\includegraphics[width=15cm]{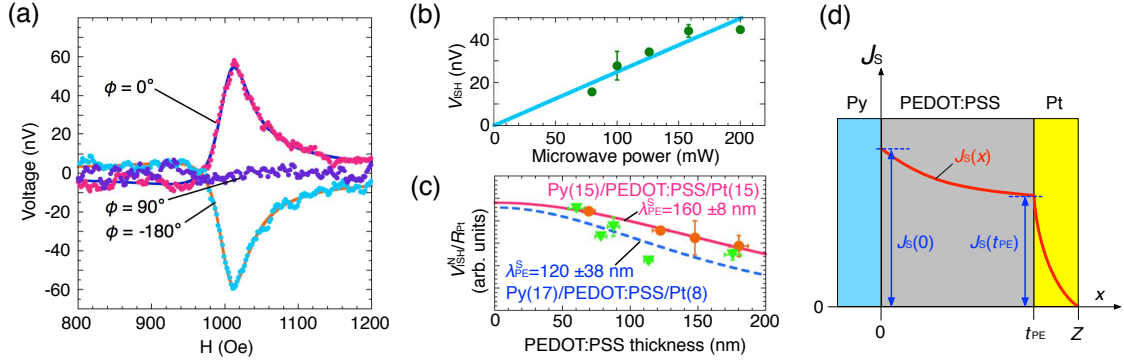}
\caption{\label{f1}(color online). (a)Magnetic field dependence of the voltage signal in Py(17)/PEDOT:PSS(60)/Pt(8) trilayer for $\phi$=0$^{\circ}$, 90$^{\circ}$, and 180$^{\circ}$. 
(b)Microwave power dependence of $V_{\rm ISH}$. 
(c)$V_{\rm ISH}^{\rm N}/R_{\rm Pt}$ as a function of PEDOT:PSS thickness. 
Solid triangles and circles represent the data for two series of samples. 
The solid and dashed lines are the fitting results using Eq. (2). 
(d)Schematic illustration of the decay of spin current in the present trilayer. }
\end{figure*}

Figure 1(a) shows the schematic of a device structure for our spin pumping experiment. The process of spin pumping dissipates the pure spin current into the PEDOT:PSS layer via the exchange interaction at the Py/PEDOT:PSS interface during the ferromagnetic resonance (FMR) excitation \cite{silsbee, mizukami, tserkovnyak, saitoh}. The transmitted spin current across the PEDOT:PSS layer is converted to the orthogonal electric field via the inverse spin Hall (ISH) effect in Pt \cite{saitoh, valenzuela, kimura}. As a result, we can detect the pure spin current through the PEDOT:PSS layer as a voltage along the Pt layer [Figure 1(b)]. The upper panel of Figure 1(c) shows the FMR spectra of Py strip in the Py(17)/PEDOT:PSS(60)/Pt(8) trilayer device. A single resonance line is observed and well fitted to the first derivative of the symmetric Lorenz function. The lower panel of Figure 1c shows the voltage signal from the Pt layer for $\phi=0^{\circ}$. The solid line is a curve fit to the sum of symmetric and asymmetric Lorentz function: 
\begin{eqnarray}
V(H)=\frac{V_{\rm S}(\Delta H/2)^2}{[(H-H_0)^2+(\Delta H/2)^2 ]} -\frac{V_{\rm A}\Delta H(H-H_0)}{[(H-H_0)^2+(\Delta H/2)^2 ]}\ ,
\end{eqnarray}
where $\Delta H$ is the spectral full width at half maximum and $H_0$ is the resonance field. $V_{\rm S}$ and $V_{\rm A}$ are symmetric and asymmetric contributions to the voltage signal, respectively \cite{saitoh}. The obtained line width and the resonance field are identical to those of FMR spectra, meaning that the voltage signal originates from the FMR of Py layer. In the present sample structure, two large contributions can be considered to generate the dc voltage signal induced by the FMR. One is the ISH voltage ($V_{\rm ISH}$) generated along the Pt layer and the other is the voltage induced by the anisotropic magnetoresistance (AMR) effect ($V_{\rm AMR}$) generated in the Py layer \cite{mecking, harder}. In the spin pumping experiment, the injected spin current takes maximum at $H_0$. Consequently, $V_{\rm ISH}$ can only contribute to $V_{\rm S}$, while $V_{\rm AMR}$ can contribute to both $V_{\rm S}$ and $V_{\rm A}$. The origin of $V_{\rm AMR}$ is interaction between the high frequency electrical current and the magnetization in the Py strip, and the in-plane magnetic field dependence of $V_{\rm AMR}$ is well understood. In this study, we have used a long rectangular Py strip as a spin injector where the dominant component of high frequency current is parallel to the long direction, and in such a case, $V_{\rm AMR}\propto \rm{sin} 2\phi$ and thus vanishes when $\phi=0^{\circ}$, 90$^{\circ}$, and 180$^{\circ}$ \cite{mecking, harder}. On the other hand, in the case of ISH effect, the conversion relation between spin current and electric field is expressed as ${\bm V_{\rm ISH}}\propto {\bm J_{\rm S}}\times {\bm \sigma} \propto {\rm cos}\phi$ where ${\bm J_{\rm S}}$ is the spin current, and ${\bm \sigma}$ is the spin polarized vector. Therefore, $V_{\rm S}$ for $\phi=0^{\circ}$ only arises from the ISH effect: $V_{\rm S}(0^{\circ})=V_{\rm ISH}(0^{\circ})$ \cite{azevedo, bai}. However, the asymmetric voltage contribution still remains for $\phi=0^{\circ}$. The origin of this component is not clear at the moment, but we consider that other magneto-transport phenomena e.g., anomalous Hall effect, can contribute to $V_{\rm A}$ \cite{chen}.

Figure 2(a) shows the magnetic field dependence of the voltage signal for $\phi=0^{\circ}$, 90$^{\circ}$, and 180$^{\circ}$. As shown in the figure, $V_{\rm S}$ changes its sign depending on the field direction, and vanishes when $\phi=90^{\circ}$. Moreover, the magnitude of $V_{\rm ISH}$ is proportional to the microwave power injected into the EPR cavity [Figure 2(b)]. Here, we take the average of $V_{\rm S}$ for $\phi=0^{\circ}$ and $180^{\circ}$ as $V_{\rm ISH}$: $V_{\rm ISH}=[V_{\rm S}(0^{\circ})-V_{\rm S} (180^{\circ})]/2$. Such tendencies are consistent with the expected behavior of $V_{\rm ISH}$ induced by the spin pumping \cite{mosendz, ando2}. The contribution of $V_{\rm ISH}$ originating from the PEDOT:PSS layer \cite{ando} is expected to be quite small in the present sample, and cannot explain the observed $V_{\rm ISH}$ in Figure 2(a) (see supplementary material). The observed $V_{\rm ISH}$ in the Pt layer is related to the spin current at the PEDOT:PSS/Pt interface, which is equivalent to the transmitted spin current through the PEDOT:PSS layer. Therefore, we can estimate the SDL of the PEDOT:PSS from the PEDOT:PSS thickness ($t_{\rm PE}$) dependence of $V_{\rm ISH}$. The plot of normalized $V_{\rm ISH}$ over resistance of Pt layer ($V_{\rm ISH}^{\rm N}/R_{\rm Pt}$) for several $t_{\rm PE}$ is shown in Figure 2(c). To consider the decay of the spin current with $t_{\rm PE}$, we have calculated the one-dimensional diffusion equation for trilayer structure with no interface resistance (see supplementary material). Based on our analysis, the spin current at the PEDOT:PSS/Pt interface [$=J_{\rm S}(t_{\rm PE})\propto V_{\rm ISH}^{\rm N}/R_{\rm Pt}$] is obtained as

\begin{eqnarray}
J_{\rm S}(t_{\rm PE})\approx J_{\rm S}(0){\rm exp}(t_{\rm PE}/\lambda_{\rm PE}^{\rm S})[1-{\rm tanh}(t_{\rm PE}/\lambda_{\rm PE}^{\rm S})]
\end{eqnarray}

for $\rho_{\rm Pt}\ll \rho_{\rm PE}^\bot$. \\
Here, $J_{\rm S}(0)$, $\lambda_{\rm PE}^{\rm S}$, $\rho_{\rm Pt}$, and $\rho_{\rm PE}^\bot$ are spin current at $x=0$ [see Figure 2(d)], SDL of PEDOT:PSS, resistivities of Pt and PEDOT:PSS along the out-of-plane direction, respectively. In the present case, $\rho_{\rm Pt}\ll \rho_{\rm PE}^\bot$ is reasonable because their present values are $\rho_{\rm Pt}=22\pm5$ $\mu \Omega$cm and $\rho_{\rm PE}^\bot=1.0\pm0.4$ k$\Omega$cm, respectively. The decay of the spin current is schematically illustrated by the solid line in Figure 2(d). The SDL of PEDOT:PSS can be estimated by fitting the data in Figure 2(c) to Equation (2), and the obtained SDLs for two series of samples are $160\pm8$ nm and $120\pm38$ nm. The difference in the SDLs for two distinct sample sets is probably arising from the quality of PEDOT:PSS films. The SDL of $\sim$140 nm on average for PEDOT:PSS obtained from our experiments is rather long compared with the SDL of 21 - 30 nm reported in the previous study \cite{ando}. Therefore, $\tau_{\rm S}$ of our PEDOT:PSS film is expected to be longer than the previous estimation of 5 - 10 $\mu$s \cite{ando}. In order to examine this expectation, we have carried out EPR measurements, in which the spin lifetime (or relaxation time) can be directly estimated.

The inset of Figure 3(a) shows an EPR spectrum for a thick ($t_{\rm PE}=10$ $\mu$m) PEDOT:PSS film at room temperature. The spectrum is well fitted to the first derivative of a single Lorenz function (solid line). In this case, the full width at the half maximum ($\Delta H_{\rm EPR}$) is related to the spin-spin relaxation (or dephasing) time $T_2$ ($\Delta H_{\rm EPR}=2/\gamma T_2$, $\gamma$: gyromagnetic ratio) \cite{poole}, and the present result ($\Delta H_{\rm EPR}=24$ Oe) corresponds to $T_2=4.7$ ns. However, the spin lifetime $\tau_{\rm S}$ discussed for dc spin current is the spin-lattice (or energy) relaxation time $T_1$, which is generally longer than $T_2$. A method to estimate $T_1$ is measuring the saturation behavior of EPR intensity ($I_{\rm EPR}$) with the microwave magnetic field ($h_{\rm ac}$). The main panel of Figure 3(a) shows the $h_{\rm ac}$ dependence of $I_{\rm EPR}$. As shown in the figure, $I_{\rm EPR}$ behaves almost linear dependence with $h_{\rm ac}$ and does not saturate up to the highest $h_{\rm ac}$. We also show the simulated results of the saturation curve with $h_{\rm ac}$ in Figure 3(b). The comparison between these two figures suggests that $T_1$ is in the range of 5 - 100 ns (the lower limit of $T_1$ is determined by $T_1=T_2$). This value is much shorter than the previously estimated $\tau_{\rm S}$ of 5 - 10 $\mu$s at room temperature \cite{ando}. In the previous study, $\tau_{\rm S}$ is indirectly estimated using the relation between SDL and $D_{\rm S}$: $\tau_{\rm S}=(\lambda^{\rm S})^2/D_{\rm S}$ with assumption of the Einstein relation (ER) for non-degenerate semiconductors, which is expressed as $D_{\rm S}=\mu k_{\rm B}T/e$ with mobility $\mu$, the Boltzmann constant $k_{\rm B}$, and the elementary charge $e$. The large discrepancy of $\tau_{\rm S}$ between our experiment and the previous estimation suggests that the estimation of $D_{\rm S}$ using the ER for non-degenerate semiconductors is not applicable to the PEDOT:PSS film. Indeed, the ER to determine $D_{\rm S}$ has different forms depending on the conduction mechanism: for thermally excited transport (non-degenerate case), $D_{\rm S}=\mu k_{\rm B}T/e$ is applicable, but for highly doped semiconductors (degenerate case), $D_{\rm S}$ is inversely proportional to the resistivity ($\rho$) and the density of states at the Fermi level $[N(E_{\rm F})]$ as similar to metallic systems, i.e., $D_{\rm S}=[e^2 N(E_{\rm F})\rho]^{-1}$. Then, we have measured the temperature dependence of electrical resistivity to discuss the conduction mechanism of PEDOT:PSS films. As shown in Figure 3(c), the out-of-plane resistivity ($\rho_{\rm PE}^\bot$) shows insulating behavior below room temperature, and the logarithm of $\rho_{\rm PE}^\bot$ is almost linear to $T^{-1/4}$ , i.e., $\rho_{\rm PE}^\bot \propto {\rm exp}(T_0/T)^{1/4}$. This is the characteristic behavior of the three-dimensional variable range hopping (3D-VRH) conduction \cite{mott, shklovskii}. In the VRH conduction, the electron transport is not dominated by thermally excited charge carriers but by tunneling between metallic localized states. Indeed, the characteristic temperature $T_0$ is expressed as $\beta/[k_{\rm B} N(E_{\rm F})\xi^3]$ with constant $N(E_{\rm F})$, where $\beta$ and $\xi$ are respectively the numerical factor ($\beta=18.1$ for 3D case) and the localization length. This temperature-independent $N(E_{\rm F})$ is the characteristic of degenerated systems: the ER for degenerated systems is applicable for the case of VRH conduction \cite{paascha}. The localization length  can be obtained from the analysis of current-voltage ($I$-$V$) characteristics along the perpendicular direction (see supplementary material for detail), and then $N(E_{\rm F})$ can be calculated from $T_0$ and $\xi$. We have measured three distinct samples, and obtained following values on average: $\rho_{\rm PE}^\bot=1.0\pm0.4$ k$\Omega$cm, $N(E_{\rm F})=8.8\pm7\times10^{17}$ eV$^{-1}$cm$^{-3}$, and $\xi=11\pm4$ nm. These values are reasonably consistent with the previous study \cite{nardes}. If we substitute $\rho$ and $N(E_{\rm F})$ by the present values to the ER for degenerate systems, $D_{\rm S}$ is estimated to be $7.1\times10^{-7}$ m$^2$/s. This value leads to the SDLs in the range of 59 - 270 nm for $T_1=5$ - 100 ns, which are comparable to the experimentally obtained SDL of $\sim$140 nm on average.

\begin{figure}
\includegraphics[width=7cm]{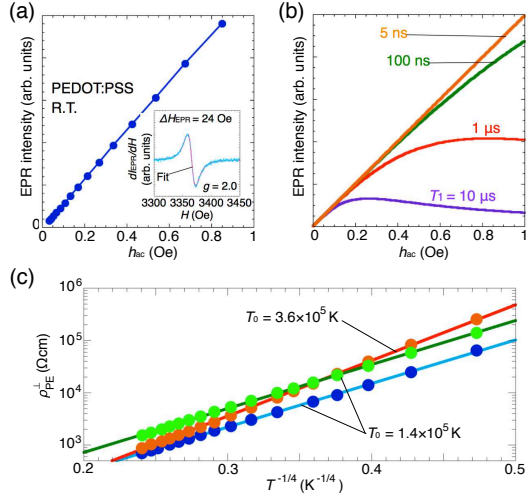}
\caption{\label{f1}(color online). 
(a)Microwave magnetic field ($h_{\rm ac}$) dependence of EPR intensity of thick PEDOT:PSS film at room temperature. 
$h_{\rm ac}$ is calculated ffrom the quality factor of the cavity. 
The intensity is obtained from the fitting as shown in the inset. 
(inset)The EPR spectra for $h_{\rm ac}=0.042$ Oe. 
The observed spectrum is well fitted to a single Lorenz function with $\Delta H_{\rm EPR}$ of 24 Oe. 
(b) The simulated behavior of the EPR intensity as a function of $h_{\rm ac}$ for several $T_1$.
Here, EPR intensity is calculated by $I_{\rm EPR}$=$h_{\rm ac}/\{1+h_{\rm ac}^2\gamma^2T_1T_2\}$ with the fixed $T_2$ (=4.7 ns).\cite{poole} 
(c) Temperature dependence of $\rho_{\rm PE}^{\bot}$ plotted with $T^{-1/4}$. 
We have measured three distinct samples. 
The solid lines are fitting based on the 3D-VRH.}
\end{figure}

\begin{figure}
\includegraphics[width=8cm]{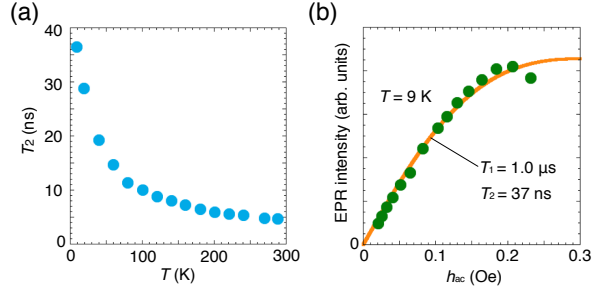}
\caption{\label{f1}(color online). (a)Temperature dependence of $T_2$ evaluated from $\Delta H_{\rm EPR}$. 
(b)$h_{\rm ac}$ dependence of EPR intensity at 9 K. 
The solid line shows the saturation curve for $T_1$=1.0 $\mu$s and $T_2$=37 ns.  }
\end{figure}

Next, we discuss the temperature dependence of spin relaxation. In OSCs, the dominant spin relaxation mechanism is still controversial, but two major candidates have been proposed, i.e., hyperfine (HF) interactions and SO couplings \cite{bobbert, yu, harmon}. Figure 4(a) shows the temperature dependence of $T_2$ evaluated from $H_{\rm EPR}(T)$. In this figure, $T_2$ gradually increase as temperature decreases, and reaches 37 ns at 9 K. Moreover, the saturation behavior of $I_{\rm EPR}$ at 9 K corresponds to $T_1 \approx 1$ $\mu$s [Figure 4(b)]. These results indicate that $\tau_{\rm S}$ of PEDOT:PSS increases at low temperatures. As discussed in ref. \cite{mizoguchi}, this result clearly indicates that the spin relaxation is arising from the spin-lattice relaxation due to the SO coupling between the polymer back born and the conduction electrons. The effect of the HF interaction would not be important in the present case. Usually, the HF coupling in organic materials is weak, and the order of the HF field is considered to be a few tens oersted \cite{naber}. Therefore, the spin relaxation (or dephasing) originating from the HF field is suppressed in sufficiently high magnetic fields ($> \sim$1000 Oe in the present experiments). The spin relaxation due to the HF interaction, on the other hand, should be discussed in the experiment at very low magnetic field and the material whose SO coupling is weak \cite{nguyen2}.

\begin{figure}
\includegraphics[width=5cm]{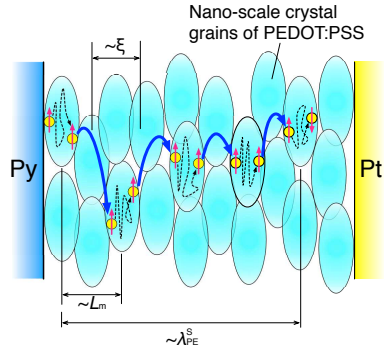}
\caption{\label{f1}(color online). Schematic of the spin transport mechanism in PEDOT:PSS film. 
The three characteristic lengths ($\xi$, $L_{\rm m}$, and $\lambda^{\rm S}_{\rm PE}$) are also shown. 
The spin relaxation mainly occurs within the PEDOT:PSS core-shell structure. 
In other words, the spin angular momentum is almost preserved in the hopping events. }
\end{figure}

In the VRH conduction, electrons hop between the adjacent hopping sites with the characteristic hopping probability. The average distance between two hopping sites is defined as the hopping length $L_{\rm m}$, which can be estimated from the $I$-$V$ characteristics and obtained as $25\pm8$ nm at room temperature (see supplementary material). At sufficiently low temperatures compared with $T_0$, the hopping probability is suppressed, so that electrons are almost trapped within the localization states during their spin lifetime. The schematic of the spin transport mechanism in PEDOT:PSS film is illustrated in Figure 5. Our experiments revealed that $\lambda_{\rm PE}^{\rm S}$ is longer than $L_{\rm m}$, suggesting that the spin angular momentum is almost preserved in the hopping event. On the other hand, the spin relaxation mainly occurs within the trapping time at the localization state.

Finally, we mention some future prospects to enhance SDLs of OSCs. The present study suggests that the SO coupling is the dominant spin relaxation process even in the organic materials. In PEDOT:PSS, the largest SO contribution is likely arising from the coupling with sulfur atoms, that is the heaviest atom in the thiophene framework. Hence, using light elements as a molecular building block is a straightforward way to achieve longer $\tau_{\rm S}$. Another direction would be reducing the doping concentration within the degenerate regime. Considerable enhancement of $D_{\rm S}$ from the low doping limit to the degenerate regime is theoretically predicted \cite{GER}. This enhancement is explained by the crossover of $D_{\rm S}$ from the thermally excited carriers to the intrinsic degenerate carriers. Therefore, the existence of intrinsic charge carriers is quite important to realize large $D_{\rm S}$. Based on the ER for degenerate systems, large $D_{\rm S}$ can be obtained by reducing $N(E_{\rm F})$, i.e., doping concentration. On the other hand, $\tau_{\rm S}$ also depends on the doping concentration. The in-situ EPR experiment with the doping concentration shows that the EPR line width decreases as the doping concentration is lowered \cite{zykwinska}. This indicates that $\tau_{\rm S}$ is elongated by reducing the doping. These discussions are qualitatively consistent with the difference of SDLs between the present and the previous studies: the present SDL of $\sim$140 nm on average is rather long compared with the previously reported value of 21 - 30 nm \cite{ando}. This discrepancy is probably due to the difference of doping concentration in PEDOT:PSS: the material used in the previous report is additionally doped with dimethyl sulfoxide solvents.

\section{Conclusions}

We have performed dynamical pure spin current transport, EPR, and charge transport experiments in a highly doped organic polymer PEDOT:PSS film. From these systematic studies, we have independently determined SDL, $\tau_{\rm S}$, and, $D_{\rm S}$. $\tau_{\rm S}$ estimated from the EPR experiment is much shorter than the previous expectation beyond the experimental ambiguity. The obtained $D_{\rm S}$ from the charge transport measurement is reasonably explained by the hopping transport mechanism in degenerate semiconductors. The SDL obtained at room temperature is of the order of one hundred nanometers, which is indeed comparable to the calculated value by using $D_{\rm S}$ and $\tau_{\rm S}$ estimated from the charge transport and EPR experiments. The most important point in this study is the comprehensive explanation of the SDL of PEDOT:PSS. The comparison between SDL and hopping length indicates that the spin angular momentum is almost preserved in the hopping event. On the other hand, this means that the spin relaxation mainly occurs in the PEDOT:PSS crystalline grains. In addition, the temperature dependence of EPR experiment shows that the main spin relaxation mechanism of PEDOT:PSS is due to the spin-lattice relaxation caused by the SO coupling. These conclusions will contribute to the full understanding of the pure spin current transport in organic semiconductors.

\section*{Acknowledgements}
The authors acknowledge T. Kato, H. Tanaka, T. Sasaki, Y. Honma, S. Watanebe, K. Furukawa for helpful discussions. 
The authors also acknowledge R. Takahashi and M. Lippmaa for the use of a surface profile measurement system. 
This work was partly supported by Grant-in-Aid for Challenging Exploratory Research (No. 24654100).



\bibliography{apssamp}

\end{document}